# Systems analysis of the phase diagrams of complex structures

B.R. Gadjiev

*International University of Nature, Society and Mans "Dubna",
141980, 19 Universitetskaya str., Dubna, Moscow Region, Russia
gadjiev@uni-dubna.ru*

**Abstract**

In frameworks of the phenomenological approach we analyze of the phase diagram of mixed compounds. We obtain space groups of symmetry of the real structures as result of phase transition from close-packed degenerate structure. The theory of a sequence of phase transitions with an incommensurate phase is constructed, and influence of an electronic subsystem and distortions on phase transitions is investigated. We investigate of dependence of the temperature interval existence of an incommensurate phase on the electron concentration and the component deformation tensor in doped semiconductors. The comparison of theoretical results with optical, dielectric and structural data is carried out and parameters of thermodynamic potential are determined.

**Introduction**

Last decades the research of the problems cultivation crystals with the given properties stimulated wide study of crystals with variable composition. Neutron and X-ray diffraction research of crystals with variable composition has revealed the basic feature of the phase diagram of the doped compounds. Thus with variation of substitutional atom concentration and external fields in doped structures the sets of modification of the structure are discovered. The detailed study of the phase diagrams of compounds with variable composition, in particular manganites or ferroelectrics, has shown, that in them there is a sequence of phase transitions. Besides, with the increase of substitutional atom concentration the temperature width of the existence of an incommensurate phase is narrowed.

In the ideal crystal the equilibrium positions of the atoms are defined by the vector $\vec{R} = n_1\vec{a}_1 + n_2\vec{a}_2 + n_3\vec{a}_3 + \vec{r}_s$, where $n_1\vec{a}_1 + n_2\vec{a}_2 + n_3\vec{a}_3$ runs on the sites of the basic lattice, and $\vec{r}_s$ ($s = 1,...,S$) sets basis in an elementary cell, and the structure has one of 230 space groups. For definition of the order parameter symmetry in doped structures it is necessary to notice, that, strictly speaking, in contrast to the ideal crystals distributions of atoms have no space group of symmetry. Namely, the substitution of the atoms in the structure breaks the periodicity of the ideal crystals.

Therefore for the definition of the order parameter symmetry in doped structures it is necessary to notice that, actually, the symmetry of doped structures is lower, than the symmetry of the similar structures, constructed by the identical spheres. However in theoretical consideration it is possible to introduce a geometrical characteristic — the order parameter, which describes this or that concrete lowering of the symmetry from the close-packed degenerated structure to the really observable derivative structure. While discussing theoretically the spontaneous symmetry breaking in doped crystals we should have it in mind that the approximation preservation of the space group of the structure, requires at least an account of the interaction of major order parameter with a secondary one, namely with deformation. Besides, the account of the energy, contributed by the substitution of atoms of the structure, is necessary. Let's note, that the results of research of genesis of structure and magnetic ordering in crystals with variable composition are presented in [1]. In according to [1] the wave vector of the magnetic ordering of the structure deviates from the initial symmetrical position under the influence of the field. This variation continuously occurs along the definite direction in Brillouin zone for incommensurate superstructures with a wave vector of general position. For the wave vectors, having the maximum symmetry of the structure, the variation of the wave vector of the superstructure has a threshold character.

In contrast to [1] in this paper we consider the case when in ends points of an interval $x \in [0,1]$ the structure with variable composition has different space groups of symmetry. In particular, we shall discuss the phase diagram of compounds with variable composition, which with change of concentration doped atoms in ends points of an interval $x \in [0,1]$ have space groups of symmetry $C_{2h}^6$ and $D_{4h}^{18}$. Dependence the temperature width of a incommensurate phase on pressure and concentration of an electronic subsystem is discussed.

### Genesis of structure of compounds $TlMeX_{2(1-m)}X'_{2m}$

From the standpoint of the genesis of the structure with variable composition, it should be noted that in the general case the symmetry of complex compounds whose structure is derived from a close-packed structure having one of the space groups $D_{4h}^6$, $O_h^5$, $O_h^9$, for the latent phase, is in reality lower than the symmetry of analogous structures made up of identical spheres. In a theoretical treatment one can always introduce a purely geometric characteristic – an order parameter - that describes some particular lowering of the symmetry. We stress that by definition the order parameter is a purely crystallographic concept that enables one to describe the symmetry difference between the probability density of the charge distribution in the degenerate and derivative structures.

The real phase of compounds $TlMeX^{\bullet}_{2(1-x)}X_{2x}$ (Me-Tl, In, Fe; X-S, Se) with space group of symmetry $C_2^3$, $C_{2h}^6$ и $D_{4h}^{18}$ can arise as a result of phase transition from latent phase with symmetry space group $O_h^9$ [2,3,4,5]. Here the transitions from latent phase to the corresponding Landau phases are described by order parameter that transforms according to the six-dimensional irreducible representation $\tau_5$ of the space group $O_h^9$, belonging to the point $\vec{k}_{10} = \frac{\pi}{2\tau}(1,1,1)$ [6,7]. The solutions for the order parameter and corresponding space group of symmetry the Landau phases has the forms $D_{4h}^{18}(0,0,c,0,0,-c)$; $C_{2h}^6(c_1,c_1,c_2,0,0,0)$, $(0,c_1,c_2,0,0,0),(c_1,c_1,c_2,c_1,c_1,c_2)$; $C_2^3(c_1,c_2,c_3,c_2,c_1,c_3),(c_1,c_2,c_3,-c_2,-c_1,-c_3)$, $(0,c_1,c_2,0,c_3,c_4)$. In the commensurate phase the volume of primitive cell is quadrupled. Besides, the irreducible representation of space group $C_{2h}^6$, corresponding to a wave vector $\vec{q}_0 = (0,0,1/4)$, describes a sequence of phase transitions $C_{2h}^6 \Rightarrow P_{1\bar{1}}^{C2/m} \Rightarrow C_2^3$. In this case at transition from high symmetry phase in a commensurate phase volume of a primitive cell quadrupled. Hence, as against phase transition $O_h^9 \Rightarrow C_2^3$, connected with irreducible representation $\tau_5$, as a result of which volume of a primitive cell quadrupled, at a sequence of phase transitions $O_h^9 \Rightarrow C_{2h}^6 \Rightarrow P_{1\bar{1}}^{C2/m} \Rightarrow C_2^3$ volume of a primitive cell is increased 16 times.

Using the condition of phase's stability, it is possible to determine the equation of line of coexistence of phases with different symmetry. In semiconductors the line of coexistence of phases is function of concentration of doped in structures atoms. Besides, in semiconductors with variable composition the interaction of an electronic subsystem with lattice has features. In fact, in the doped crystals is present distortion of structure. In semiconductors it can produce impurity levels or localization of electrons and, hence to influence concentration dependence of parameters of thermodynamic potential. Besides, since in semiconductors the electron concentration on impurity levels is function of temperature that, hence, the line of coexistence of phases also will be function of temperature [8,9].

**Influence of an electronic subsystem on phase transitions in semiconductors**
$TlMeX_{2(1-m)}X'_{2m}$

According to symmetry arguments the thermodynamic potential functional of the ferroelectrics-semiconductors of the TlMeX$_2$ type with space group of symmetry $C_{2h}^6$ is presented by expression

$$F = \frac{1}{d}\int_0^d f(z)dz, \quad f = f_{lat} + mf_{elek}, \tag{1}$$

where $m$ — electron concentration on the impurity, $f_{elek}$ — energy of the electrons on the impurity levels[8,9].

In accordance with the condition of invariance, the lattice part of thermodynamic potential is expressed as:

$$f_{lat}(z) = \frac{\alpha_0}{2}\rho^2 + \frac{\beta_0}{4}\rho^4 + \gamma_0\rho^8\cos 8\varphi - \delta_0\rho^2\varphi' + \frac{k_0}{2}\rho^2(\varphi')^2, \tag{2}$$

here $\rho$ and $\varphi$ — amplitude, and phase of the order parameter, accordingly and $\alpha_0 = \alpha_{00}(T - T_0)$, $\beta_0, \gamma_0, \delta_0$; $k_0$ —parameters of expansion of the thermodynamic potential functional, $T_0$ — temperature of phase transition from high symmetry phase in a incommensurate phase in ideal crystal.

Near the phase transition the energy corresponding to electron on the impurity levels can be expressed by the order parameter as:

$$f_{elek} = E_0 + \frac{a}{2}\rho^2 + \frac{b}{4}\rho^4 + \Gamma\rho^8\cos 8\varphi - \Delta\rho^2\varphi' + \frac{k'}{2}\rho^2(\varphi')^2. \tag{3}$$

Here - $a, b, \Gamma, \Delta, k', \lambda'$ — coefficients of expansion of thermodynamic potential of the order parameter.

Thus, thermodynamic potential functional is possible to present as:

$$F = \frac{1}{d}\int_0^d f(z)dz \tag{4}$$

$$f(z) = \frac{\alpha}{2}\rho^2 + \frac{\beta}{4}\rho^4 + \gamma\rho^8\cos 8\varphi - \delta\rho^2\varphi' + \frac{k}{2}\rho^2(\varphi')^2, \tag{5}$$

where $\alpha = \alpha_0 + ma$, $\beta = \beta_0 + bm$, $\gamma = \gamma_0 + m\Gamma$, $\delta = \delta_0 + m\Delta$, $k = k_0 + mk'$.

Therefore, the expansion coefficients of the thermodynamic potential depend on electron concentration on impurity levels [1,9,10].

The harmonic solution representing an incommensurate phase is $\varphi = \kappa x$, and from the condition of minimum of thermodynamic potential we find $\kappa_0 = \frac{\delta}{k}$.

In this case there is a shift of critical temperature, which is represented by expression

$$T_{ic} = T_I - \frac{1}{\alpha_{00}}ma = T_0 + \frac{1}{\alpha_{00}}(\frac{\delta^2}{k} - ma) \tag{6}$$

The concentration dependence of the temperature width of a incommensurate phase is deduced from expression

$$4\rho^3 = \pi(\frac{\delta^2}{\gamma k})^{\frac{1}{2}}\frac{\kappa}{E(\kappa)},$$

$$\rho^3 = (-\frac{\alpha^*}{\beta})^{\frac{3}{2}}, \tag{7}$$

where $0 \leq \tau \leq 1$.

With the account $\alpha^* = \alpha_{00}(T - T_{ic})$ and $\tau \to 1$ (therefore $T \to T_c$), for temperature width of a incommensurate phase is obtained

$$\frac{\beta_0 + bm}{\alpha_{00}} \left( \frac{\pi^2 (\delta_0 + m\Delta)^2}{16(k_0 + mk')(\gamma_0 + m\Gamma)} \right)^{\frac{1}{3}} - \frac{a}{\alpha_{00}} m = (T_I - T_c) \tag{8}$$

Neglecting dependence of parameters $\beta$, $\delta$, $k$ и $\gamma$ from concentration it is possible to determine critical value of concentration, at which temperature width of a incommensurate phase becomes equal to zero:

$$m_c = \frac{\beta}{a} \left( \frac{\pi^2 \delta^2}{16k\gamma} \right)^{\frac{1}{3}}.$$

From the equation (8) follows, that generally $(T_I - T_c)$ is nonlinear function of concentration and depends on a combination of parameters of thermodynamic potential functional. With increase concentration $m$ the temperature width of a incommensurate phase is narrowed.

**Influence of deformation on phase transitions in semiconductors $TlMeX_{2(1-m)}X'_{2m}$**

For the description of phase transitions in structures with variable composition is necessary to take into account influence a component of displacement tensors.

In accordance with symmetry arguments thermodynamic potential functional is possible to present as:

$$F = \frac{1}{d} \int_0^d f(z) dz, \quad f(z) = f_{e-l} + f_{el}, \tag{9}$$

where

$$f_{t-l}(z) = \frac{\alpha}{2} \rho^2 + \frac{\beta}{4} \rho^4 + \gamma \rho^8 \cos 8\varphi - \delta \rho^2 \varphi' + \frac{k}{2} \rho^2 (\varphi')^2,$$

$$f_{el} = -\delta_1 u_{xx} \rho^2 \frac{\partial \varphi}{\partial x} + \frac{\lambda}{2} u_{xx}^2 + \varsigma u_{xx} \rho^2 - \sigma_{xx} u_{xx}.$$

After minimization $f(z)$ we find:

$$u_{xx}^0 = \frac{1}{\lambda} \left( \delta_1 \rho^2 \frac{\partial \varphi}{\partial x} + \sigma_{xx} - \varsigma \rho^2 \right) \tag{10}$$

With the account the equation (10) in the equation (9) we obtain:

$$f(z) = f_{e-l} + f_{el} = \left( \frac{\alpha}{2} - \frac{\varsigma \sigma_{xx}}{\lambda} \right) \rho^2 + \left( \frac{\beta}{4} - \frac{\varsigma^2}{2\lambda} \right) \rho^4 + $$
$$+ \gamma \rho^8 \cos 8\varphi - \left( \delta \rho^2 + \frac{\delta_1}{\lambda} \rho^2 \sigma_{xx} - \frac{\varsigma \delta_1}{\lambda} \rho^4 \right) \varphi' + \left( \frac{k}{2} \rho^2 - \frac{\delta_1^2}{2\lambda} \rho^4 \right) (\varphi')^2 - \frac{\sigma_{xx}^2}{\lambda} \tag{11}$$

From the condition of a minimum of thermodynamic potential we find value of a wave vector of modulation

$$\kappa_{0\sigma} = \frac{\delta + \frac{\delta_1}{\lambda} \sigma_{xx} - \frac{\varsigma \delta_1}{\lambda} \rho^2}{k - \frac{\delta_1^2}{\lambda} \rho^2} \approx \frac{\delta}{k} + \frac{\delta_1}{k\lambda} \sigma_{xx}, \tag{12}$$

which is linear function $\sigma_{xx}$.

The shift critical temperatures is determined from the equation

$$\frac{\alpha}{2} - \frac{\varsigma \sigma_{xx}}{\lambda} - \frac{1}{2k} \left( \delta + \frac{\delta_1}{\lambda} \sigma_{xx} \right)^2 = 0. \tag{13}$$

With the account, that, $T_I = T_0 + \dfrac{1}{\alpha_{00}} \dfrac{\delta^2}{k}$, for the shift of critical temperature is obtained:

$$T_{ic}^{\sigma} = T_I - \dfrac{1}{\alpha_{00}} ma + \dfrac{1}{\alpha_{00}} (\dfrac{2\varsigma \sigma_{xx}}{\lambda} + \dfrac{1}{k} \dfrac{\delta \delta_1}{\lambda} \sigma_{xx} + \dfrac{1}{2k} \dfrac{\delta_1^2}{\lambda^2} \sigma_{xx}^2) \tag{14}$$

In this case the temperature width of a incommensurate phase is represented as:

$$\dfrac{\beta_0 + bm - \dfrac{2\varsigma^2}{\lambda}}{\alpha_{00}} (\dfrac{\pi^2 (\delta_0 + m\Delta + \dfrac{\delta_1}{\lambda} \rho^2 \sigma_{xx})^2}{16(k_0 + mk')(\gamma_0 + m\Gamma)})^{\tfrac{1}{3}} - \dfrac{a}{\alpha_{00}} m + \tag{15}$$

$$+ \dfrac{1}{\alpha_{00}} (\dfrac{2\varsigma \sigma_{xx}}{\lambda} + \dfrac{1}{k} \dfrac{\delta \delta_1}{\lambda} \sigma_{xx}) = (T_I - T_c),$$

and therefore depends on the elastic tensors component. If with increase $m$ the temperature width of an incommensurate phase goes to zero, with increase $\sigma_{xx}$ $(T_I - T_c)$ increased.

Let's consider the case, when because of strong distortion of structure the electronic subsystem is located, and, hence, the parameters of thermodynamic potential do not depend on electron concentration.

In this case electron concentration in equation (15) can be neglected, and then temperature width of a incommensurate phase is defined as

$$\dfrac{\beta_0 - \dfrac{2\varsigma^2}{\lambda}}{\alpha_{00}} (\dfrac{\pi^2 (\delta_0 + \dfrac{\delta_1}{\lambda} \rho^2 \sigma_{xx})^2}{16(k_0)(\gamma_0)})^{\tfrac{1}{3}} + \dfrac{1}{\alpha_{00}} (\dfrac{2\varsigma \sigma_{xx}}{\lambda} + \dfrac{1}{k} \dfrac{\delta \delta_1}{\lambda} \sigma_{xx}) = (T_I - T_c) \tag{16}$$

Represents the special interest, when the substitution of structure results also in strong distortion of structure, to such that last can cause nonlinear effects. In this case the thermodynamic potential presented as

$$f_{t-l}(z) = \dfrac{\alpha}{2} \rho^2 + \dfrac{\beta}{4} \rho^4 + \gamma \rho^8 \cos 8\varphi - \delta \rho^2 \varphi' + \dfrac{k}{2} \rho^2 (\varphi')^2,$$

$$f_{el} = -\delta_1 u_{xx} \rho^2 \dfrac{\partial \varphi}{\partial x} + \dfrac{\lambda}{2} u_{xx}^2 + \dfrac{\nu}{4} u_{xx}^4 - \sigma_{xx} u_{xx} \tag{17}$$

Minimization thermodynamic potential (16) lead to the cubic equation for definition of $u_{xx}$:

$$\nu u^3 + \lambda u - \delta_1 \rho^2 \dfrac{\partial \varphi}{\partial x} - \sigma = 0 \tag{18}$$

The sign of quantity $D = (\dfrac{\lambda}{2\nu})^3 - (\dfrac{\delta_1 \rho^2 \dfrac{\partial \varphi}{\partial x} + \sigma}{2\nu})^2$ defines the number of real solutions of last equation. In particular, if $D > 0$ the equation (17) has one real solution, whereas if $D < 0$ the equation (17) has three real solutions.

Hence, account of nonlinear effects on $u$ results to splitting the phase diagram of physical system. The phases differ with equilibrium values of quantity $u$.

Discussion

The space group of symmetry of crystals $Tl$ is $O_h^9$. Therefore we assume, that in degenerate case the $Tl$ based crystals have volume-centered cubic structure, from which in result substitution of structure arise really observable phases of compounds $TlMeX_{2(1-m)}X'_{2m}$ (Me-Tl, In, Fe; X-S, Se). Six dimensional irreducible representation $\tau_5$ of group $O_h^9$, corresponding to the point $\vec{k}_{10}$, lead to phase transition to the phases with space groups of symmetry $D_{4h}^{18}$, $C_{2h}^6$ and $C_2^3$. Irreducible representation of space group $C_{2h}^6$, corresponding to the wave $\vec{q}_0 = (0,0,1/4)$, describes a sequence of phase transitions high symmetry- incommensurate- commensurate phase as $C_{2h}^6 \Rightarrow P_{1\bar{1}}^{C2/m} \Rightarrow C_2^3$ and at transition from high symmetry phase in a commensurate one volume of a primitive cell quadrupled. In semiconductors and doped structures temperature width of an incommensurate phase is function of charges concentration and of component displacement tensor. The structural researches show that is higher then 235.1 $^0C$ is stable $\beta - Tl$ with volume centered cubic lattice. The structures of layered compounds $TlGaS_2$, $TlGaSe_2$ и $TlInS_2$ are monoclinic and at room temperature have space group of symmetry $C_{2h}^6$. The structure of chained compounds $TlInSe_2$ и $TlInTe_2$ is tetragonal and at room temperature have space group of symmetry $D_{4h}^{18}$. The crystals $TlFeS_2$ и $TlFeSe_2$ in high symmetry phase have a space group of symmetry $C_2^3$.

Besides the incommensurate phase in $TlGaSe_2$ и $TlInS_2$ crystals exists in an interval of temperatures from~120K up to 110K and from ~216K up to ~197K, accordingly [6,7]. The space group of symmetry of high-symmetry phase of $TlGaSe_2$ и $TlInS_2$ crystals is $C_{2h}^6$. A wave vector of modulation of the incommensurate phase is $\vec{q}_{inc} = (\sigma,0,0,0.25)$, where σ = 0.04 [8].

The compounds $TlInS_{2(1-x)}Se_{2x}$ at change of concentration in an interval $x \in [0,1]$ are undergone concentration phase transition $D_{4h}^{18}(x=1) \to C_{2h}^6(x=0)$. To critical value of concentration $x = 0.4$ corresponds a compound $TlInS_{2(1-x)}Se_{2x}$, temperature width incommensurate phase of which is equal to zero. The concentration dependences of temperature width of an incommensurate phase of compounds $TlInS_{2(1-x)}Se_{2x}$, determined from optical and of dielectric measurements, are presented in [11]. Using equation (16) and experimental results of work [11] we from fitting procedure have determined combination of parameters of thermodynamic potential as:

$$\frac{\beta_0 - \frac{2\varsigma^2}{\lambda}}{\alpha_{00}} (\frac{\pi^2\delta_0^2}{16k_0\gamma_0})^{\frac{1}{3}} = 16.6, \frac{1}{\alpha_{00}}(\frac{2\varsigma}{\lambda} + \frac{1}{k}\frac{\delta\delta_1}{\lambda}) = -164.1,$$

$$\frac{\beta_0 - \frac{2\varsigma^2}{\lambda}}{\alpha_{00}} (\frac{\pi^2\delta_0^2}{16k_0\gamma_0})^{\frac{1}{3}} = 12.4, \frac{1}{\alpha_{00}}(\frac{2\varsigma}{\lambda} + \frac{1}{k}\frac{\delta\delta_1}{\lambda}) = -30.1.$$

The obtained result shows, that at substitution of structure by atoms of greater radius (replacement $S \Rightarrow Se$) lead to in narrowing temperature width of a incommensurate phase, if this replacement is arise as negative pressure.

Using equation (8) and experimental results of work [11] we from fitting procedure have determined combination of parameters of thermodynamic potential as:

$$\frac{\beta_0}{\alpha_{00}}(\frac{\pi^2(\delta_0)^2}{16(k_0)(\gamma_0)})^{\frac{1}{3}} = 12.4, \frac{a}{\alpha_{00}} = 30.1 \quad \frac{\beta_0}{\alpha_{00}}(\frac{\pi^2(\delta_0)^2}{16(k_0)(\gamma_0)})^{\frac{1}{3}} = 16.6, \frac{a}{\alpha_{00}} = 164.1.$$

Therefore the under influence of electron subsystem the temperature width of a incommensurate phase is narrowed.

We thank I. Lukyanchuk for many useful discussions.